\documentclass[12pt]{article}
\usepackage{graphicx}
\usepackage{bm}
\usepackage{amssymb,amsmath,amsfonts,mathtools,mathrsfs}
\usepackage{cite}
\usepackage[colorlinks=true,linkcolor=blue,citecolor=blue,urlcolor=blue]{hyperref}

\newcommand{\ket}[1]{\left|#1 \right\rangle}
\newcommand{\bra}[1]{\left\langle#1 \right|}
\newcommand{\braket}[2]{\langle #1| #2 \rangle}

\begin{document}

\title{Novel Variational Approach for Analysis of Photonic Crystal Slabs}

\maketitle

\author{Mohammad Hasan Aram and Sina Khorasani}

School of Electrical Engineering, Sharif University of Technology, Tehran, Iran

khorasani@sina.sharif.edu

\begin{abstract}
We propose a new method based on variational principle for analysis of photonic crystal (PC) slabs. Most of the methods used today treat PC slab as a three-dimensional (3D) crystal and this makes them very time and/or memory consuming. In this method we use Bloch theorem to expand the field on infinite plane waves which their amplitudes depend on the component perpendicular to the slab surface. By approximating these amplitudes with appropriate functions, we can find modes of PC slabs almost as fast as we can find modes of a two-dimensional (2D) crystal. Besides this advantage , we can also calculate radiation modes with this method which is not feasible with 3D Plane Wave Expansion (PWE)  method.
\end{abstract}


%
%
%
%
%

\section{Introduction}

Since the prediction of forbidden gaps in the band-structure of some periodic dielectric structures that we now call photonic crystals (PCs) \cite{Yablo}, these structures have affected many scientific fields and have found many industrial applications. Today, PCs are used to increase solar cells efficiency \cite{Zhou,Mutitu,Zeng,Bermel}, improve lasers and light emmiting diodes specs \cite{Hirayama,Englund,Ichikawa}, and design new kind of waveguides and optical fibers \cite{Chutinan,Lin,Johnson,Loncar,Cerqueira}. These nano-structures are a good platform for optical circuits \cite{McNab,Marin} and seem to find application in some new research fields such as quantum information and quantum computation \cite{Akahane,Ohta,Thon,Englund2,Brossarda,Lvovsky}.

Periodicity of PCs dielectric constant can be in one, two, or three dimensions \cite{Gmitter,Ho,KRAUSS}. Fabrication of full 3D PCs or perfect 2D ones is difficult or sometimes impossible. This fact has made scientist to do their best to work with slab of PCs because they are fabricated easily with conventional fabrication technologies. PC slabs have periodic permittivity in two dimensions, but finite thickness along the third one. They are sometimes known as finite thickness 2D crystals. These crystals can confine and guide electromagnetic waves according to distributed Bragg reflection in the plane of the slab and total internal reflection in the slab normal direction \cite{Johnson,Steven,Chow,Villeneuve,Noda}.  
Up to now many methods have been proposed for analysis of PCs which fall into two categories; frequency domain and time domain. Frequency domain methods such as finite element, plane wave expansion, and finite difference frequency domain deal with phasors of electromagnetic fields, but time domain ones like finite difference time domain (FDTD) find fields evolution according to Maxwell's equation. Since PC slab is not a perfect 2D crystal, we need to simulate it as a full 3D one and this has made its analysis time and/or memory consuming. Some innovative methods have been proposed to solve this problem \cite{Nekuee,Shi,Qiu,Aram,Shi2}, but almost all of them have some shortcomings among them we can name, lower accuracy and limited frequency range analysis.

In this paper we first introduce our new method which can find PC slab modes by calculating eigenvalues of a matrix twice as large as the matrix appears in 2D PWE method. Then we obtain band structure of a sample PC slab and compare it with the results of conventional 3D PWE and FDTD methods. We also compare eigen field of this crystal at a high symmetry point obtained from the proposed method with that of 3D PWE method. To show its ability to analyze PC waveguides we find guided modes of a sample PC waveguide, and finally we compare the efficiency of this method against 3D PWE method. 

\section{Variational method for PC slab analysis}

According to variational principle one can finds an estimation to smallest eigenvalues of a Hermitian operator $\mathbb{L}$, by choosing a suitable trial function as the eigen function of the operator and then trying to minimize 
\[ \frac{\bra{f} \mathbb{L} \ket{f}}{\braket{f}{f}}, \]
where $\ket{f}$ is the trial function.
In fact the popular PWE method is based on the variational principle. In this method the trial function, that is one of the electric or magnetic fields, is chosen according to Bloch theorem to be
\[ F (\mathbf{r}) = e^ {-j \boldsymbol\kappa \cdot \mathbf{r}} \Phi_{\boldsymbol\kappa} (\mathbf{r}) = \sum_{\mathbf{G}} \varphi_{\mathbf{G}} e^{-j (\boldsymbol\kappa + \mathbf{G}) \cdot \mathbf{r} }, \]
where $\Phi_{\boldsymbol\kappa} (\mathbf{r})$ is a periodic function with the same periodicity as that of the dielectric constant of the crystal, $\bm{\kappa}$ is the Bloch wave vector and $\mathbf{G}$ equals $\sum_{i=1}^3 m_i \mathbf{b}_i$, with $\mathbf{b}_i$s being the primitive vectors of the reciprocal lattice. The second equality is written by substituting $\Phi_{\boldsymbol\kappa} (\mathbf{r})$ with its Fourier series expansion.
In photonic crystal slab there is no periodicity in the vertical direction of the slab surface. Hence to use the standard PWE method we need to create an artificial periodicity in this direction. This means that we are estimating the field dependency on the normal component of the slab by a Fourier series. Actually this is not a good estimation, because the number of Fourier series coefficients that should be determined can be very large.

\begin{figure}[h]
\centering
\includegraphics[width=\linewidth]{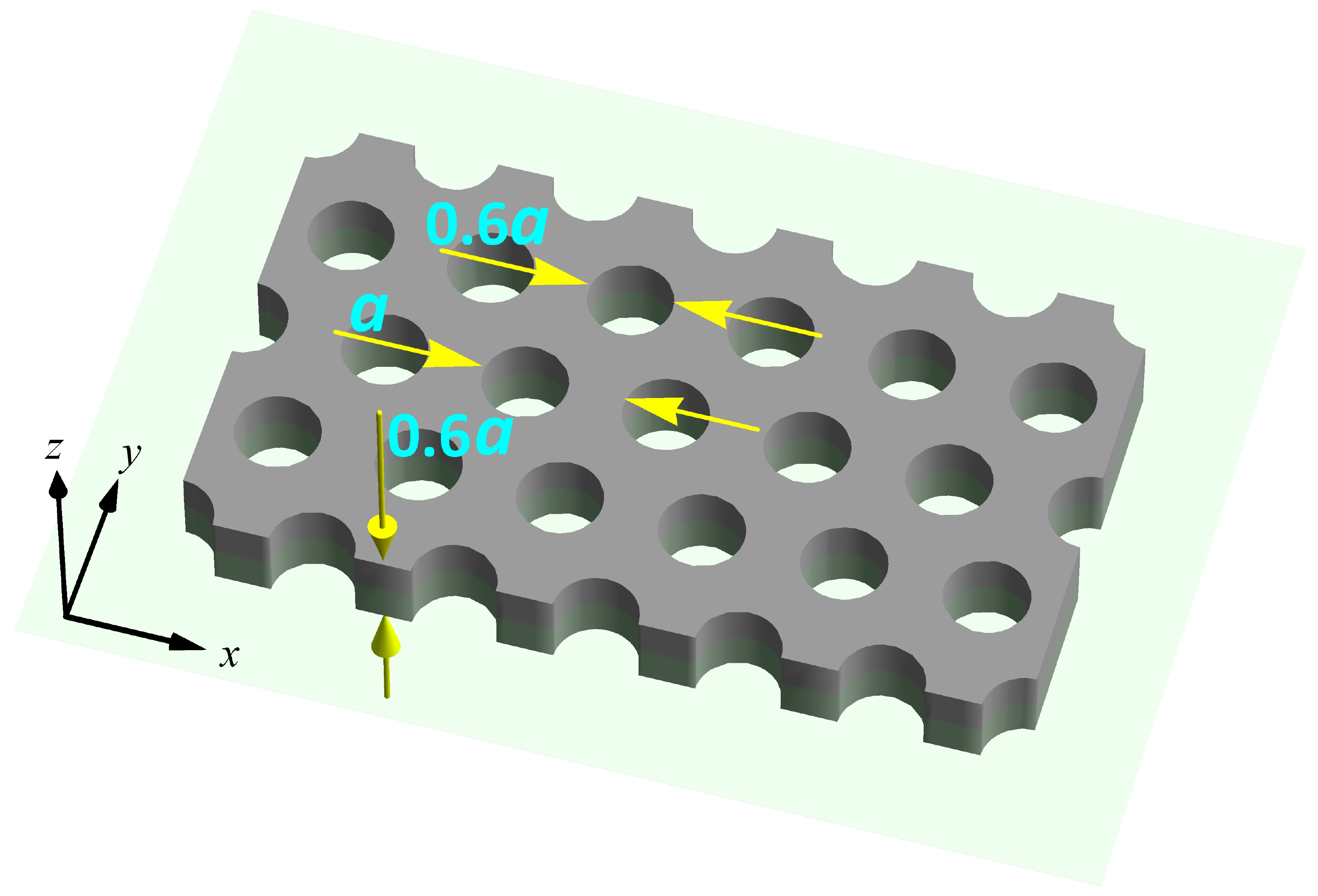}
\caption{\label{f1} PC slab analyzed in this paper constructed by etching circular air columns through a dielectric slab with dielectric constant $\epsilon_r=11.9$}
\end{figure}

Here we want to show we can choose a simpler trial function for the field dependency on the normal component of the slab which results in faster calculation of crystal modes. For clarity of the formulation written in the remaining of the paper, suppose we want to analyze the photonic crystal slab shown in figure~\ref{f1}. This crystal is composed of a triangular lattice of circular air columns etched through a dielectric slab with dielectric constant $\epsilon_r=11.9$, and thickness $t=0.6a$, where $a$ is the lattice constant. The radius of air columns equals $0.3a$. 
According to Bloch theorem we can write magnetic field as
\begin{equation} \label{e1}
\mathbf{H}(\mathbf{r}) = e^{-j \boldsymbol\kappa \cdot \mathbf{r}_{xy}} \Phi(\mathbf{r}_{xy} ,z) ,  
\end{equation}
where $\Phi(\mathbf{r}_{xy} ,z)$ is a periodic function of $x$ and $y$ for every value of $z$. If we substitute $\Phi(\mathbf{r}_{xy} ,z)$ by its Fourier series, we can write
\begin{align}
\mathbf{H}(\mathbf{r}) 
&= \sum_{\mathbf{G}} \mathbf{h}_{\mathbf{G}}(z) e^{-j (\bm{\kappa} + \mathbf{G}) \cdot \mathbf{r}_{xy}}  \nonumber \\
&= \sum_{\mathbf{G}} \left( h_{\mathbf{G}_{\|}}(z) \mathbf{e}_{\mathbf{G}_{\|}} + h_{\mathbf{G}_{\perp}}(z) \mathbf{e}_{\mathbf{G}_{\perp}}+ h_{\mathbf{G}_{z}}(z) \hat{z} \right) \nonumber \\ 
& \qquad \quad \times  e^{-j (\boldsymbol\kappa + \mathbf{G}) \cdot \mathbf{r}_{xy}} \label{e2},  
\end{align}
where $\mathbf{e}_{\mathbf{G}_{\|}}$ and $\mathbf{e}_{\mathbf{G}_{\perp}}$ are unit vectors along and perpendicular to $\bm{\kappa} + \mathbf{G}$ respectively.
Here $\mathbf{G}$ is the vector in reciprocal lattice of a 2D crystal with the same pattern as that of our PC slab.
To find a good estimation for $h_{\mathbf{G}_{\|}}(z)$, $h_{\mathbf{G}_{\perp}}(z)$, and $h_{\mathbf{G}_{z}}(z)$ we need to survey field distribution in a dielectric slab waveguide problem. We know in that problem, amplitudes of guided electric and magnetic fields inside the slab are sinusoidal functions of $z$ component, but decay exponentially outside the slab as $|z|$ increases, that is  
\begin{align}
F_i(\mathbf{r},t) &= F^0_i (z) e^{-j \bm{\kappa} \cdot \mathbf{r}_{xy}} e^{j \omega t} \quad , \quad  \left( i = \| , \perp , z \right) \nonumber \\
&= \left\{
\begin{array}{lcr}
A_{\text{o}}\sin(k z) \ \text{or} \ A_{\text{e}}\cos(k z) & , &|z| \le t/2\\
C_u e^{-\alpha (z-t/2)} & , &  z > t/2 \\
C_l e^{\alpha (z+t/2)} & , &  z < -t/2
\end{array} \right.  \label{e3}
\end{align}
where $F$ stands for electric $E$, or magnetic $H$, fields, $\omega$ is the angular frequency, and $A_{\text{o}}$, $A_{\text{e}}$ , $C_u$, $C_l$, $\alpha$, and $k$ are constants. It can be shown
\begin{equation}\label{e4}
\omega^2 = \frac{k^2 + \kappa^2}{\epsilon_r} \quad \text{and} \quad \alpha^2 = \kappa^2 - \omega^2, 
\end{equation}
where $ \kappa = |\bm{\kappa}|$.
Considering each of the even/odd or TE/TM modes of the waveguide we can also write four relations between $\alpha$ and $k$,
\begin{align}
\text{TE:} & \left\{
\begin{array}{l}
\alpha  =  k \tan (k t/2) \\
\alpha  = -k \cot (k t/2)
\end{array} \right. \nonumber \\
\text{TM:} &\left\{
\begin{array}{l}
\alpha  = k \tan (k t/2) / \epsilon_r \\
\alpha  = -k \cot (k t/2) / \epsilon_r
\end{array} \right. .\label{e5}
\end{align}
We can simply show that fields profile do not change outside the slab if we carve out the triangular lattice of circular columns, that is electromagnetic field decays exponentially outside the PC slab as $|z|$ increases. If we can approximate its profile inside the PC slab by sinusoidal functions, then our estimated trial function in TE-like mode becomes   
\begin{align}
h_{\mathbf{G}_{\|}}(z) &= A_{\mathbf{G}_{\|}} \left\{
\begin{array}{lcr}
\sin(k_{\mathbf{G}_{\|}} z)  & , &|z| \le t/2\\
\sin (k_{\mathbf{G}_{\|}} t/2) e^{-\alpha_{\mathbf{G}} (z-t/2)} & , &  z > t/2 \\
-\sin (k_{\mathbf{G}_{\|}} t/2) e^{\alpha_{\mathbf{G}} (z+t/2)} & , &  z < -t/2
\end{array} \right. \nonumber \\ 
&= A_{\mathbf{G}_{\|}}  f_{\mathbf{G}_{\|}}(z) \nonumber \\ 
h_{\mathbf{G}_{\perp}}(z) &= A_{\mathbf{G}_{\perp}} \left\{
\begin{array}{lcr}
\sin(k_{\mathbf{G}_{\perp}} z)  & , &|z| \le t/2\\
\sin (k_{\mathbf{G}_{\perp}} t/2) e^{-\alpha_{\mathbf{G}} (z-t/2)} & , &  z > t/2 \\
-\sin (k_{\mathbf{G}_{\perp}} t/2) e^{\alpha_{\mathbf{G}} (z+t/2)} & , &  z < -t/2
\end{array} \right. \nonumber \\
&= A_{\mathbf{G}_{\perp}}  f_{\mathbf{G}_{\perp}}(z) \nonumber \\ 
h_{\mathbf{G}_{z}}(z) &= -j |\boldsymbol\kappa + \mathbf{G}| A_{\mathbf{G}_{\|}} \nonumber \\
& \times \left\{
\begin{array}{lcr}
\cos(k_{\mathbf{G}_{\|}} z) /k_{\mathbf{G}_{\|}}  & , &|z| \le t/2\\
\frac{\sin (k_{\mathbf{G}_{\|}} t/2)}{\alpha_{\mathbf{G}}} e^{-\alpha_{\mathbf{G}} (z-t/2)} & , &  z > t/2 \\
\frac{\sin (k_{\mathbf{G}_{\|}} t/2)}{\alpha_{\mathbf{G}}} e^{\alpha_{\mathbf{G}} (z+t/2)} & , &  z < -t/2
\end{array} \right. \nonumber \\
&=A_{\mathbf{G}_{\|}} f_{\mathbf{G}_{z}}(z).  \label{e6}
\end{align}
$h_{\mathbf{G}_{\|}}(z) $ and $h_{\mathbf{G}_{\perp}}(z)$ are chosen to be continuous at $|z|=t/2$ and $h_{\mathbf{G}_{z}}(z)$ is written such that  $\nabla \cdot \mathbf{H} (\mathbf{r}) =0$. TM-like mode trial functions can be written in the same manner.

$A_{\mathbf{G}_{\|}}$, $A_{\mathbf{G}_{\perp}}$, $k_{\mathbf{G}_{\|}}$,  $k_{\mathbf{G}_{\perp}}$, and $\alpha_{\mathbf{G}}$ are parameters which
can be determined by minimizing
\begin{equation}\label{e7}
\frac{\bra{\mathbf{H}(\mathbf{r})} \mathbb{L}_{\mathbf{H}} \ket{\mathbf{H}(\mathbf{r}) }}{\braket{\mathbf{H}(\mathbf{r}) }{\mathbf{H}(\mathbf{r}) }}, 
\end{equation}
where
\[ \mathbb{L}_{\mathbf{H}} =\nabla \times \left( \frac{1}{\epsilon_r(\mathbf{r})} \nabla \times (\cdot) \right). \]
This is again a time consuming problem. To further simplify it, suppose we have suitable values for $\alpha_{\mathbf{G}}$, $k_{\mathbf{G}_{\|}}$, and $k_{\mathbf{G}_{\perp}}$. Then, minimization of expression (\ref{e7}) becomes minimizing $\bra{A_{\mathbf{G}}} \mathbb{M} \ket{A_{\mathbf{G}}}$ provided that  $\bra{A_{\mathbf{G}}} \mathbb{N} \ket{A_{\mathbf{G}}} = 1$, where
\begin{align}
&\ket{A_{\mathbf{G}}} = \begin{bmatrix}\begin{pmatrix}
A_{\mathbf{G}_{\|}} \\
A_{\mathbf{G}_{\perp}}
\end{pmatrix} \end{bmatrix}_{\mathbf{G}}  , \ 
\mathbb{M} = \begin{bmatrix} \eta_{\mathbf{G}' - \mathbf{G}}\begin{pmatrix}
m_{11} &  m_{12} \\
m_{21} &  m_{22}
\end{pmatrix}  \end{bmatrix}_{\mathbf{G},\mathbf{G}'} \nonumber \\
&\text{and} \ \mathbb{N} = \begin{bmatrix} \begin{pmatrix}
n_{11} &  0 \\
0 &  n_{22}
\end{pmatrix}  \end{bmatrix}_{\mathbf{G},\mathbf{G}'}. \label{e8}
\end{align}
In (\ref{e8}) $\eta_{\mathbf{G}}$ are Fourier series coefficients of $\eta(\mathbf{r})=1/\epsilon_r(\mathbf{r})$, and elements of matrices $\mathbb{M}$ and $\mathbb{N}$ equals
\begin{align}
m_{11} &= \mathbf{e}_{\mathbf{G}_{\|}} \cdot \mathbf{e}_{\mathbf{G}'_{\|}} \Big( | \boldsymbol\kappa + \mathbf{G}|^2 \braket{f'_{\mathbf{G}'_{\|}}(z)}{f_{\mathbf{G}_z}(z)}  \nonumber\\
&+ 
| \boldsymbol\kappa + \mathbf{G}|^2 | \boldsymbol\kappa + \mathbf{G}'|^2 \braket{f_{\mathbf{G}'_z}(z)}{f_{\mathbf{G}_z}(z)} \nonumber \\
&+ \braket{f'_{\mathbf{G}'_{\|}}(z)}{f'_{\mathbf{G}'_{\|}}(z)} + | \boldsymbol\kappa + \mathbf{G}'|^2 \braket{f_{\mathbf{G}'_z}(z)}{f'_{\mathbf{G}_{\|}}(z)}  \Big) ,\nonumber \\
m_{12} &= \mathbf{e}_{\mathbf{G}_{\perp}} \cdot \mathbf{e}_{\mathbf{G}'_{\|}} \Big( \braket{f'_{\mathbf{G}'_{\|}}(z)}{f'_{\mathbf{G}_{\perp}}(z)} \nonumber \\
& \qquad \qquad \qquad + | \boldsymbol\kappa + \mathbf{G}'|^2 \braket{f_{\mathbf{G}'_z}(z)}{f'_{\mathbf{G}_{\|}}(z)} \Big) ,\nonumber \\
m_{21} &= \mathbf{e}_{\mathbf{G}_{\|}} \cdot \mathbf{e}_{\mathbf{G}'_{\perp}} \Big( \braket{f'_{\mathbf{G}'_{\perp}}(z)}{f'_{\mathbf{G}_{\|}}(z)} \nonumber \\
& \qquad \qquad \qquad + | \boldsymbol\kappa + \mathbf{G}|^2 \braket{f'_{\mathbf{G}'_{\perp}}(z)}{f_{\mathbf{G}_z}(z)} \Big) ,\nonumber \\
m_{22} &=  | \boldsymbol\kappa + \mathbf{G}| | \boldsymbol\kappa + \mathbf{G}'| \braket{f_{\mathbf{G}'_{\perp}}(z)}{f_{\mathbf{G}_{\perp}}(z)} \nonumber \\
&+ \mathbf{e}_{\mathbf{G}_{\perp}} \cdot \mathbf{e}_{\mathbf{G}'_{\perp}} \braket{f'_{\mathbf{G}'_{\perp}}(z)}{f'_{\mathbf{G}_{\perp}}(z)}, \nonumber \\
n_{11} &= \left\{
\begin{array}{lcr}
\braket{f_{\mathbf{G}_{\|}}(z)}{f_{\mathbf{G}_{\|}}(z)} \\
+ | \boldsymbol\kappa + \mathbf{G}|^2 \braket{f_{\mathbf{G}_{z}}(z)}{f_{\mathbf{G}_{z}}(z)} & , & \mathbf{G} = \mathbf{G}'\\
0 & , & \mathbf{G} \neq \mathbf{G}' 
\end{array} \right. , \nonumber \\
n_{22} &= \left\{
\begin{array}{lcr}
\braket{f_{\mathbf{G}_{\perp}}(z)}{f_{\mathbf{G}_{\perp}}(z)} & , & \mathbf{G} = \mathbf{G}'\\
0 & , & \mathbf{G} \neq \mathbf{G}' 
\end{array} \right. .  \label{e9}
\end{align}

Using Lagrange multipliers method we can determine $\ket{A_{\mathbf{G}}}$ elements by solving the generalized eigenvalue problem,
\begin{equation}\label{e10}
\mathbb{M} \ket{A_{\mathbf{G}}} + \lambda \mathbb{N} \ket{A_{\mathbf{G}}} = 0.
\end{equation}

As said above we first need to set the values of $\alpha_{\mathbf{G}}$, $k_{\mathbf{G}_{\|}}$, and $k_{\mathbf{G}_{\perp}}$ in order to obtain the simplified eigenvalue problem of (\ref{e10}). 
For this purpose we temporarily assume angular frequency $\omega$, to be that of dielectric slab waveguide with the same slab thickness and permittivity equals the effective permittivity of PC slab. We mean by effective permittivity, the coefficient of Fourier series of $\epsilon_r(\mathbf{r})$ with $\mathbf{G}=0$. Please notice that we only consider the first band of dispersion diagram of slab waveguide, that is we set
\[ \omega = \omega_1 (| \boldsymbol\kappa |) . \]
After that using (\ref{e4}) we can write
\[ \alpha_{\mathbf{G}} = \sqrt{| \boldsymbol\kappa + \mathbf{G}|^2 - \omega^2}, \]
and then $k_{\mathbf{G}_{\|}}$ and $k_{\mathbf{G}_{\perp}}$  are obtained by solving the following equations
\begin{align}
\text{TE-like:} & \left\{
\begin{array}{l}
\alpha_{\mathbf{G}}  =  k_{\mathbf{G}_{\|}} \tan ( k_{\mathbf{G}_{\|}} t/2) \\
\alpha_{\mathbf{G}}  = -k_{\mathbf{G}_{\perp}} \cot (k_{\mathbf{G}_{\perp}} t/2) / \epsilon_{r_{\text{eff}}}
\end{array} \right. \nonumber \\
\text{TM-like:} &\left\{
\begin{array}{l}
\alpha_{\mathbf{G}}  =  - k_{\mathbf{G}_{\|}} \cot ( k_{\mathbf{G}_{\|}} t/2) \\
\alpha_{\mathbf{G}}  = k_{\mathbf{G}_{\perp}} \tan (k_{\mathbf{G}_{\perp}} t/2) / \epsilon_{r_{\text{eff}}}
\end{array} \right. ,\nonumber
\end{align}
where $\epsilon_{r_{\text{eff}}}$ is the effective permittivity of PC slab.

\section{\label{s3} Results comparison}
\begin{figure}[ht!]
\includegraphics[width=\linewidth]{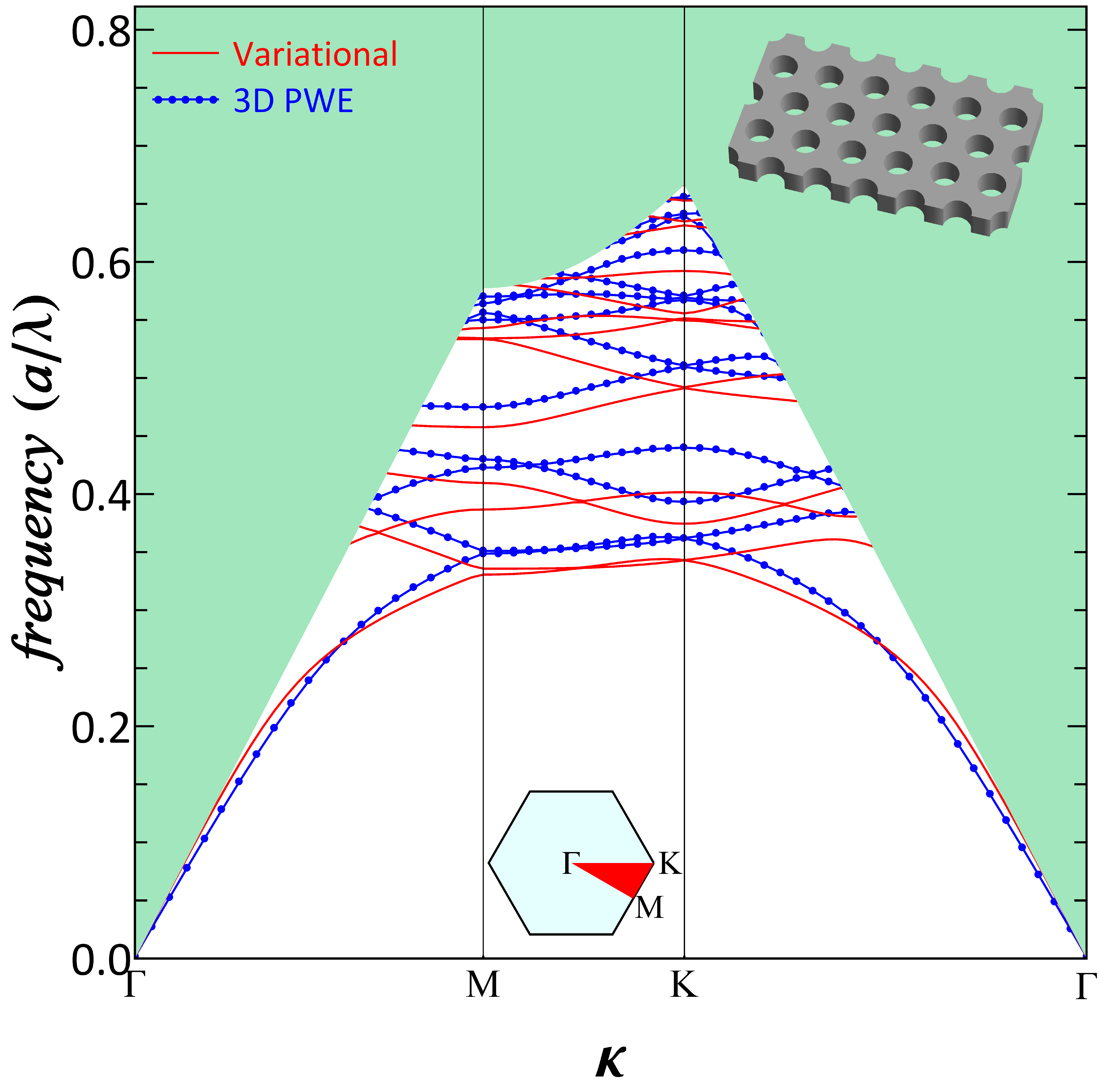}
\caption{\label{f2} Comparison of TM-like band structures of the crystal shown in figure~\ref{f1} obtained from the variational method by that of 3D PWE method. Green region shows the light cone.}
\end{figure}

\begin{figure}[ht!]
\includegraphics[width=\linewidth]{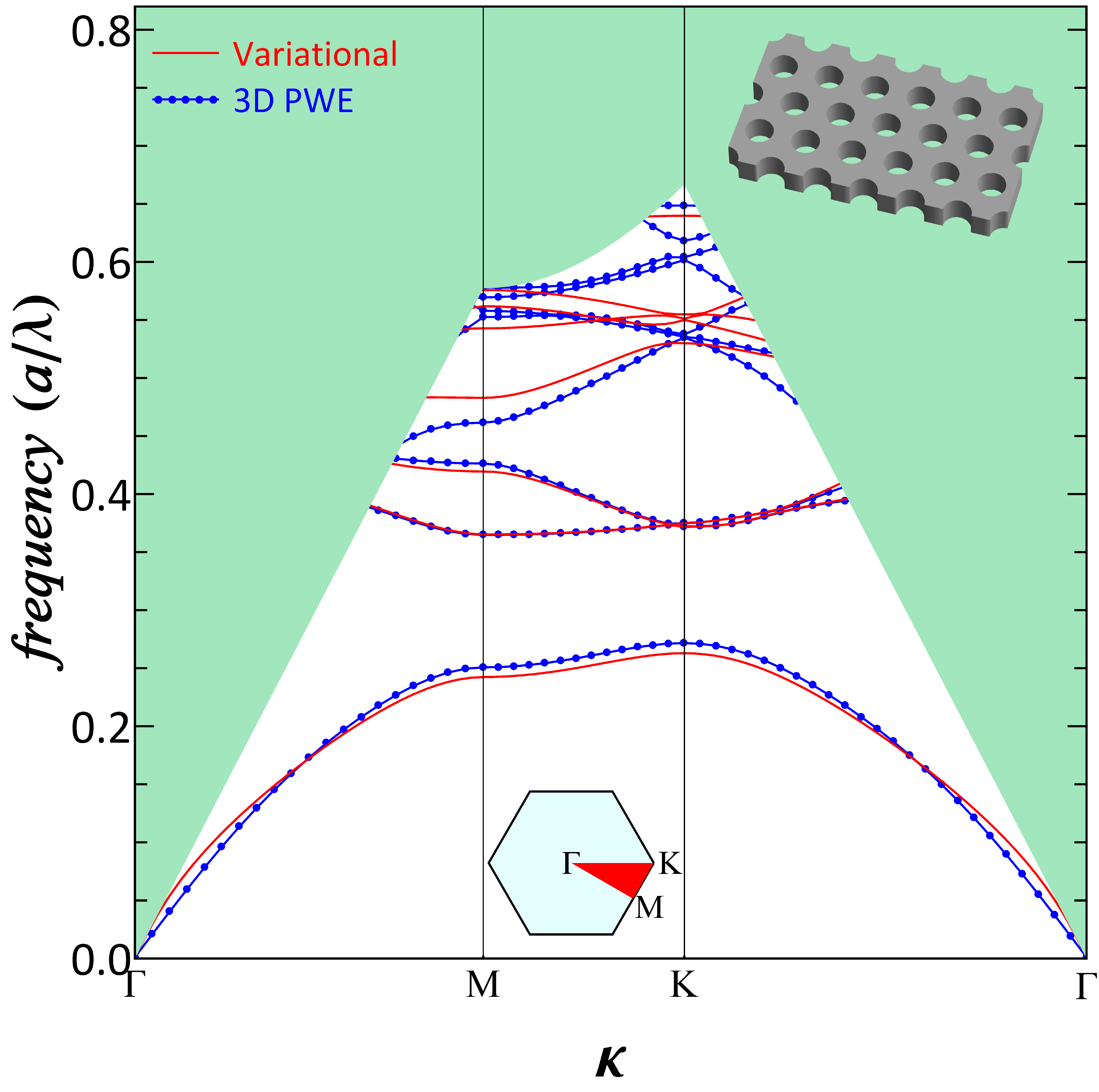}
\caption{\label{f3} Comparison of TE-like band structures of the crystal shown in figure~\ref{f1} obtained from the variational method by that of 3D PWE method}
\end{figure}

The calculated TM-like and TE-like band structures of the crystal shown in figure~\ref{f1} by the proposed variational method have been compared with the results of standard 3D PWE method in figures~\ref{f2} and \ref{f3} respectively.
For PWE method, number of Fourier series terms in $xy$ plane was bounded to $N_x=N_y=5$ and to $N_z=6$ in $z$ direction. We also used $N_x=N_y=5$  for variational method.   
As is seen results have good agreements. In fact in TM-like mode our method gives more precise result.
Discrepancy in high frequency bands of TE-like mode is mostly due to inaccuracy of 3D PWE method near the light cone. 
As mentioned, in 3D PWE method we need to make an artificial periodicity along the slab normal direction by introducing infinite slabs similar to the main one above and below it. This is done because field decay exponentially outside the slab and we can assume other slabs do not change the field distribution near and inside the main slab. But when we go near to the light cone that is when we are on the edge of confined and radiation modes of the slab, the field decay slowly outside the slab and this assumption is not valid any more so 3D PWE method loses its accuracy.

\begin{figure}[ht!]
\includegraphics[width=\linewidth]{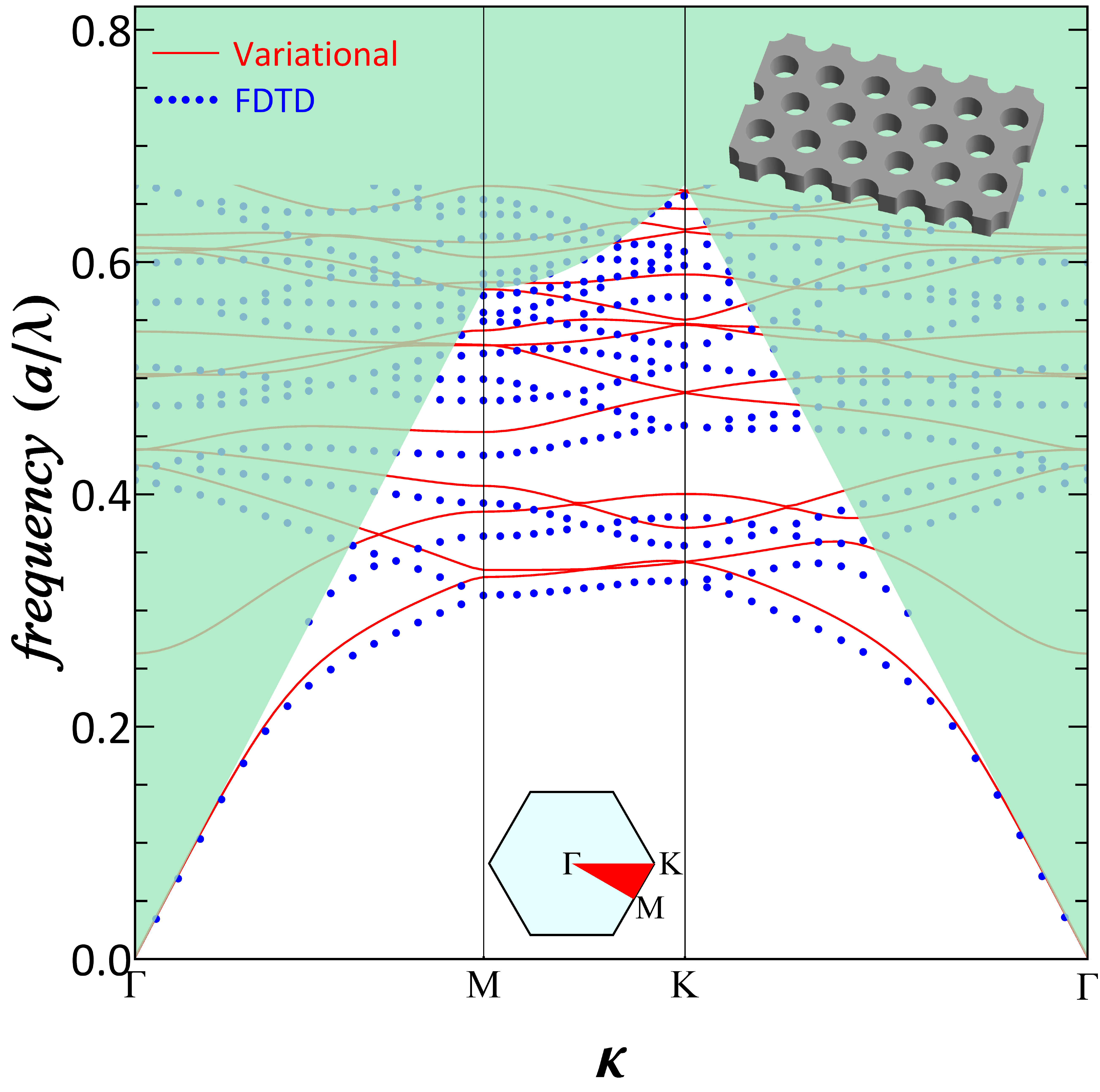}
\caption{\label{f4} Guided and radiation TM-like modes of the crystal of figure~\ref{f1} calculated by the variational method and compared with the results of FDTD method}
\end{figure}

One of the advantages of the proposed method over the 3D PWE one is that besides calculating guided modes of PC slab, we can also obtain radiation modes by this method. figure~\ref{f4} shows guided and radiation TM-like modes of the crystal of figure~\ref{f1} calculated by the variational method and  compares it with FDTD method result.

\begin{figure}[ht!]
\includegraphics[width=0.05\linewidth]{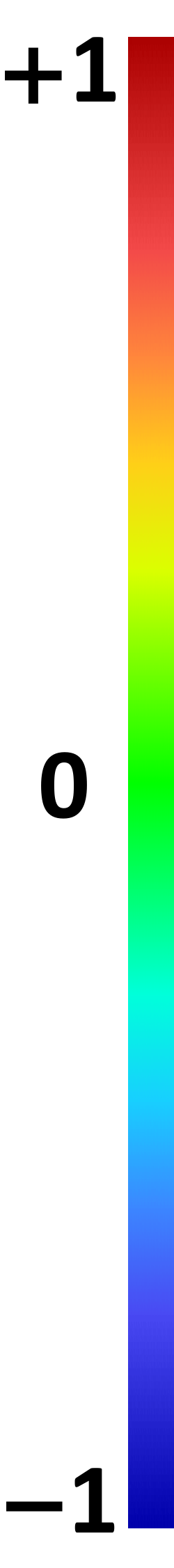} \hfill
\includegraphics[width=0.44\linewidth]{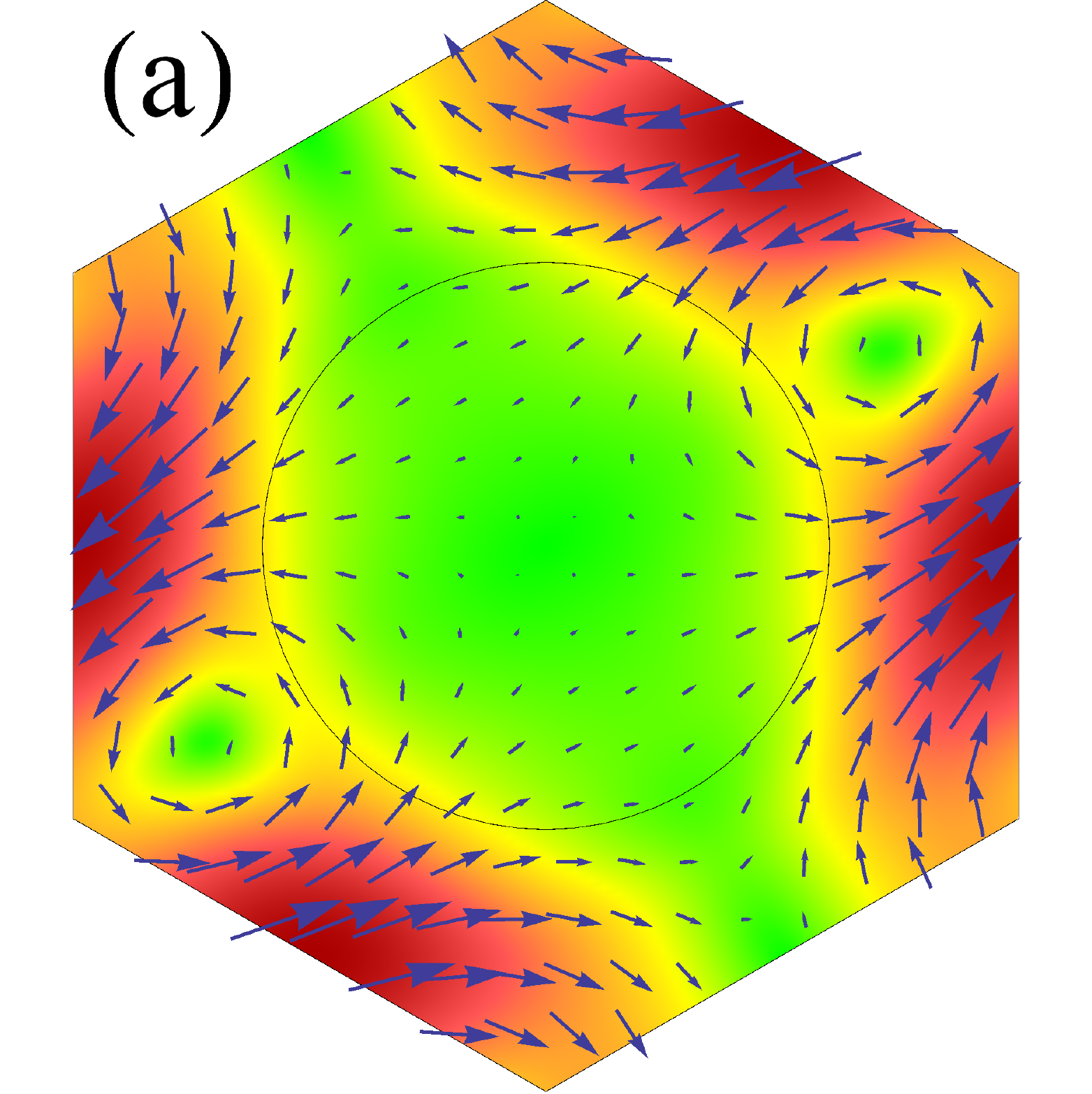} \hfill
\includegraphics[width=0.44\linewidth]{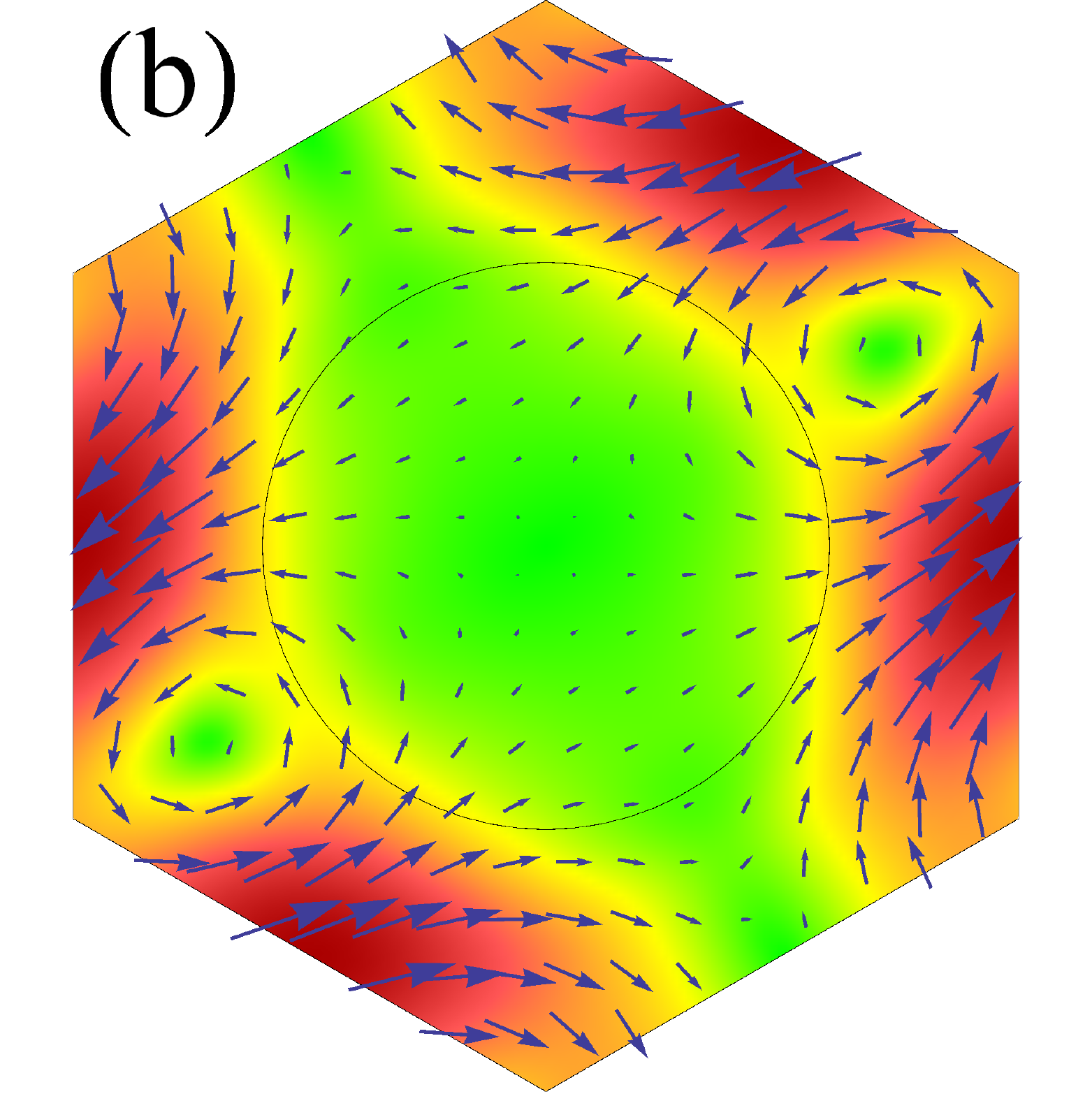}\\
\includegraphics[width=0.05\linewidth]{Fig5band} \hfill
\includegraphics[width=0.44\linewidth]{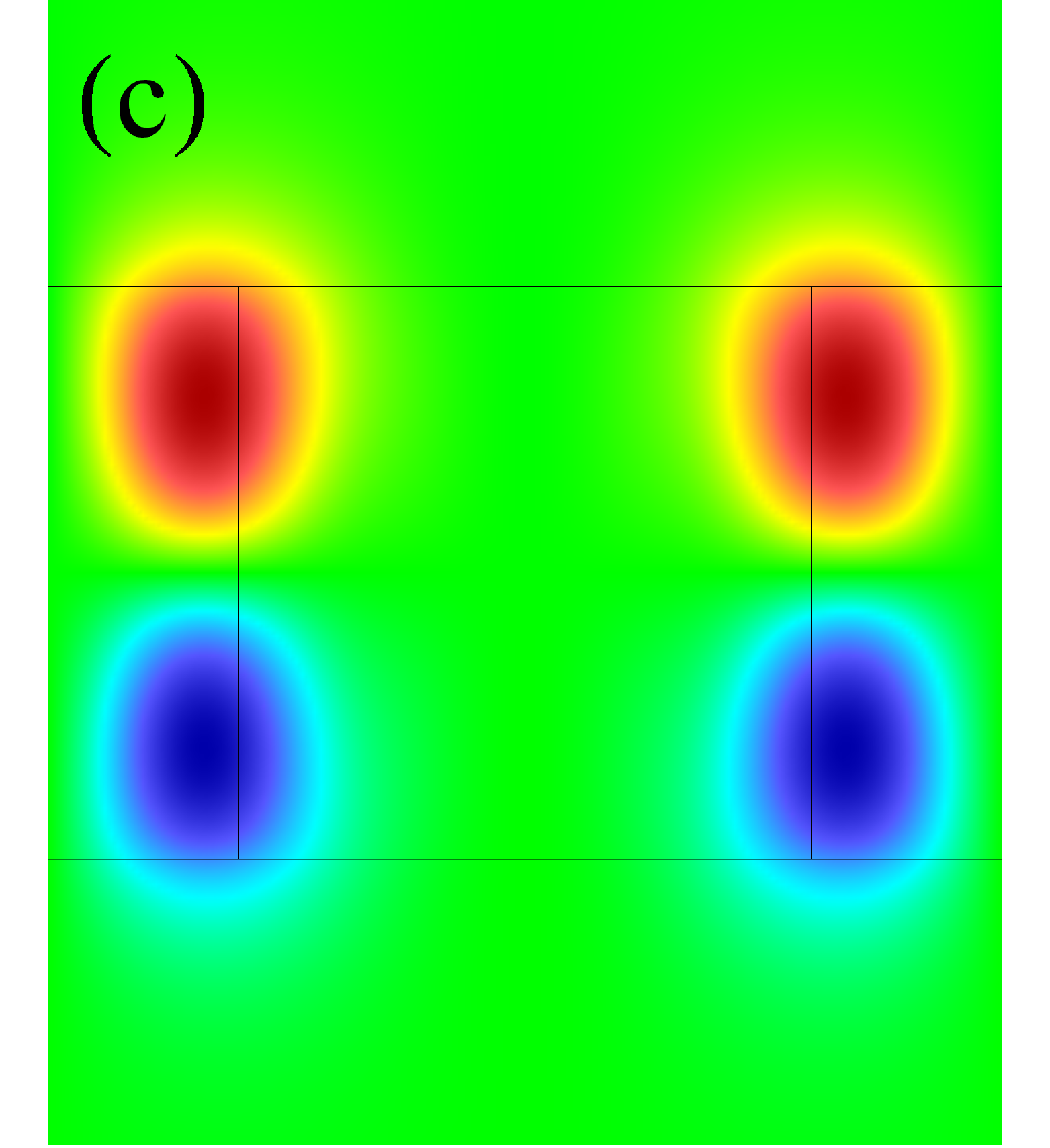}\hfill
\includegraphics[width=0.44\linewidth]{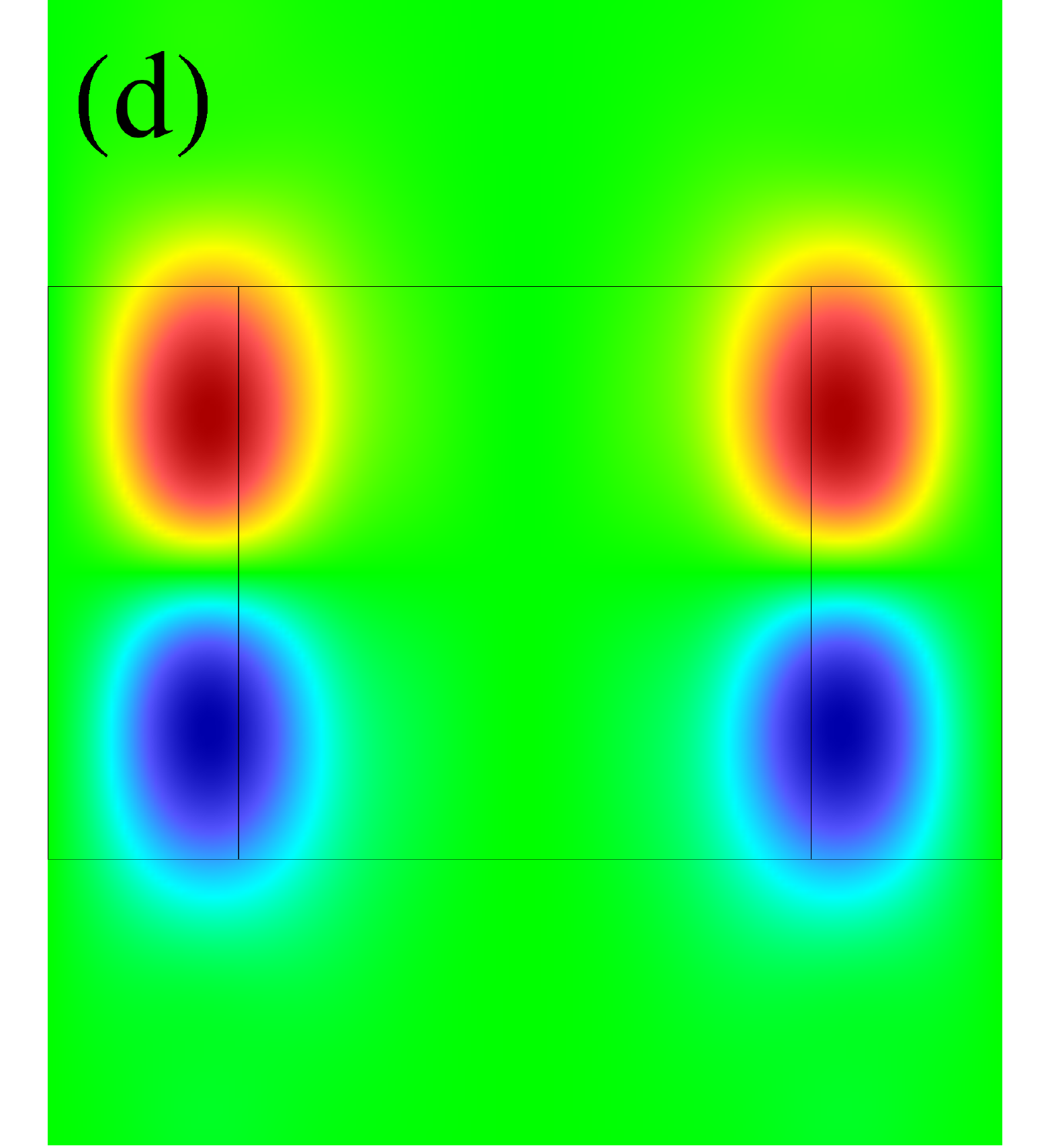}
\caption{ \label{f5} TM-like magnetic field distribution at the high-symmetry point $\text{M}^{(5)}$ in the mid-plane of the crystal of figure~\ref{f1}, obtained from (a) variational and (b) 3D PWE methods and $z$ component of the magnetic field in a vertical plane to the crystal obtained from (c) variational and (d) 3D PWE methods.
 }
\end{figure}

We have also used our method to plot TM-like magnetic field distribution in the mid-plane of the PC slab and in a vertical plane to it at the high-symmetry point of the reciprocal lattice $\text{M}^{(5)}$, fifth mode at the M point, and compared it with the result of PWE method in figure~\ref{f5}. Again we see good agreement between the results.  

\begin{figure}[ht!]
\includegraphics[width=\linewidth]{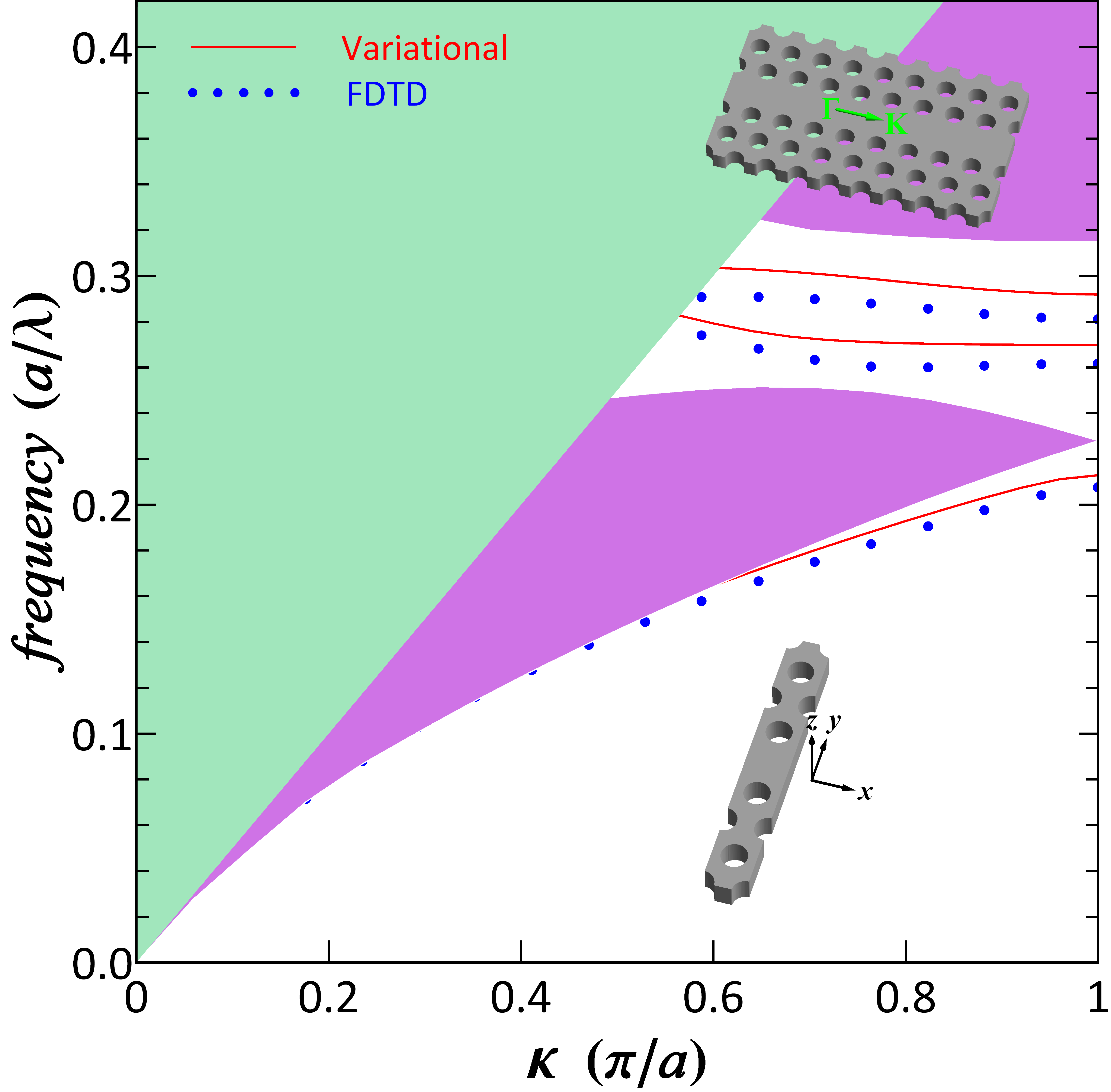}
\caption{\label{f6} Dispersion diagram of the PC slab waveguide shown in the upper inset, calculated by both FDTD and variational methods. This waveguide is built by filling one row of air columns in the crystal of figure~\ref{f1} in $\Gamma-\text{K}$ direction. The lower inset shows the unit cell used for analysis. Magenta regions represent extended modes inside the crystal.}
\end{figure}

To examine our method capability in dealing with large unit cell crystals, we plotted in figure~\ref{f6} dispersion diagram of a PC slab waveguide by both FDTD and variational methods. This waveguide is built by filling one row of air columns with dielectric in the crystal of figure~\ref{f1} in $\Gamma-\text{K}$ direction. It is shown in the upper inset of figure~\ref{f6} and has three guided modes which our method has found all of them. 
The lower inset of figure~\ref{f6} shows the unit cell used to analyze this waveguide. Number of Fourier series terms in variational method was limited to $N_x=7$ and $N_y=23$ but can get more accurate result by going beyond this limits.   

\section{\label{s4} Efficiency comparison}

\begin{figure}[h!]
\includegraphics[width=\linewidth]{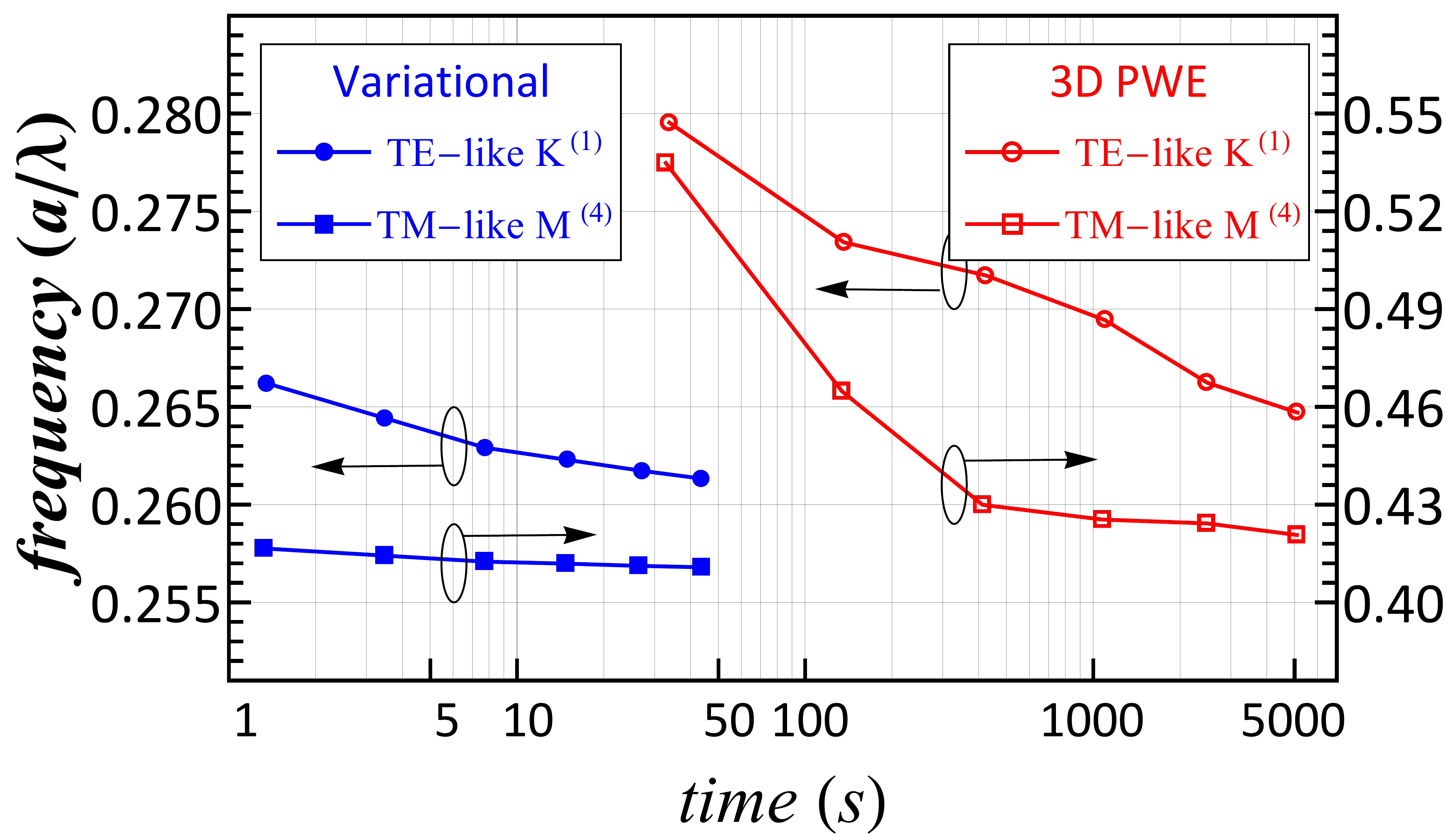}
\caption{\label{f7} Convergence time comparison between variational and PWE methods. Frequencies of the first TE-like and fourth TM-like modes at high-symmetry points M and K for the crystal of figure~\ref{f1} have been plotted versus the time it takes to calculate them. A personal computer equipped with a 32-bit $\text{Intel}^{\textregistered} \text{core}^{\text{TM}}2$ Duo CPU has been used to calculate these frequencies.}
\end{figure}

It is clear that the proposed method is much faster than the conventional 3D PWE method. The reason is that in our method we have to calculate a matrix which its dimensions are of the order $O(N^2)$, where $N$ is the number of Fourier series terms. While in 3D PWE method the matrix dimensions are of the order $O(N^3)$. To illustrate this fact quantitatively we have plotted in figure~\ref{f7}  frequencies of the first TE-like and fourth TM-like modes at high-symmetry points M and K respectively of the introduced crystal in figure~\ref{f1} versus the time it takes to calculate them. These calculations have been performed with a personal computer equipped with a 32-bit  $\text{Intel}^{\textregistered} \text{core}^{\text{TM}}2$  Duo CPU. The frequencies are calculated by variational and 3D PWE methods. Results with minimum calculation time are related to $N_x=N_y=N_z-1=3$ limitations on terms of Fourier series and ones with maximum calculation time are related to $N_x=N_y=N_z-1=8$. It can be seen that the variational method converges to the final value at least one order of magnitude faster than PWE method. 

\section{\label{s5} Conclusions }

We presented a new method for fast analysis of PC slabs. In this method which is similar to PWE method instead of creating an artificial periodicity along the normal component of the slab and then approximating the field dependency on this component by a Fourier series, we adopted field distribution in a dielectric slab waveguide to estimate field in PC slabs. Results of this fast method are in good agreements with that of other standard methods. We also compared the convergence time of our method to the final result with that of 3D PWE method. Our method shows at least one order of magnitude faster convergence.

\section*{Acknowledgement}
This work has been supported by the Research Deputy of Sharif University of Technology.

\bibliography{Ref}

\providecommand{\noopsort}[1]{}\providecommand{\singleletter}[1]{#1}%
\providecommand{\newblock}{}
\begin{thebibliography}{10}
\expandafter\ifx\csname url\endcsname\relax
  \def\url#1{{\tt #1}}\fi
\expandafter\ifx\csname urlprefix\endcsname\relax\def\urlprefix{URL }\fi
\providecommand{\eprint}[2][]{\url{#2}}

\bibitem{Yablo}
Yablonovitch E 1987 {\em Phys.\ Rev.\ Lett.\/} {\bf 58} 2059--2062

\bibitem{Zhou}
Zhou D and Biswas R 2008 {\em J.\ Appl.\ Phys.\/} {\bf 103} 093102

\bibitem{Mutitu}
Mutitu J~G, Shi S, Chen C, Creazzo T, Barnett A, Honsberg C and Prather D~W
  2008 {\em OPTICS EXPRESS\/} {\bf 16} 15238--15248

\bibitem{Zeng}
Zeng L, Bermel P, Yi Y, Alamariu B~A, Broderick K~A, Liu J, Hong C, Duan X,
  Joannopoulos J and Kimerling L~C 2008 {\em Appl.\ Phys.\ Lett.\/} {\bf 93}
  221105

\bibitem{Bermel}
Bermel P, Luo C, Zeng L, Kimerling L~C and Joannopoulos J~D 2007 {\em OPTICS
  EXPRESS\/} {\bf 15} 16986--17000

\bibitem{Hirayama}
Hirayama H, Hamano T and Aoyagi Y 1996 {\em Appl.\ Phys.\ Lett.\/} {\bf 69}
  791--793

\bibitem{Englund}
Altug H, Englund D and Vu\v{c}kovi\'{c} J 2006 {\em Nature Physics\/} {\bf 2}
  484--488

\bibitem{Ichikawa}
Ichikawa H and Baba T 2004 {\em Appl.\ Phys.\ Lett.\/} {\bf 84} 457--459

\bibitem{Chutinan}
Chutinan A and Noda S 1999 {\em Appl.\ Phys.\ Lett.\/} {\bf 75} 3739--3741

\bibitem{Lin}
Lin S~Y, Chow E, Hietala V, Villeneuve P~R and Joannopoulos J~D 1998 {\em
  Science\/} {\bf 282} 274--276

\bibitem{Johnson}
Johnson S~G, Villeneuve P~R, Fan S and Joannopoulos J~D 2000 {\em Phys.\ Rev.\
  B\/} {\bf 62} 8212--8222

\bibitem{Loncar}
Lon\v{c}ar M, Doll T, Vu\v{c}kovi\'{c} J and Scherer A 2000 {\em J.\ Lightwave\
  Technol.\/} {\bf 18} 1402--1411

\bibitem{Cerqueira}
Cerqueira A 2010 {\em Rep.\ Prog.\ Phys.\/} {\bf 73} 024401

\bibitem{McNab}
McNab S~J, Moll N and Vlasov Y~A 2003 {\em OPTICS EXPRESS\/} {\bf 11}
  2927--2939

\bibitem{Marin}
Solja\v{c}i\'{c} M, Luo C, Joannopoulos J~D and Fan S 2003 {\em OPTICS
  EXPRESS\/} {\bf 28} 637--639

\bibitem{Akahane}
Akahane Y, Asano T, Song B~S and Noda S 2003 {\em Nature\/} {\bf 425} 944--947

\bibitem{Ohta}
Ohta R, Ota Y, Nomura M, Kumagai N, Ishida S, Iwamoto S and Arakawa Y 2011 {\em
  Appl.\ Phys.\ Lett.\/} {\bf 98} 173104

\bibitem{Thon}
Thon S~M, Rakher M~T, Kim H, Gudat J, Irvine W~T~M, Petroff P~M and Bouwmeester
  D 2009 {\em Appl.\ Phys.\ Lett.\/} {\bf 94} 111115

\bibitem{Englund2}
Englund D, Majumdar A, Faraon A, Toishi M, Stoltz N, Petroff P and
  Vu\v{c}kovi\'{c} J 2010 {\em Phys.\ Rev.\ Lett.\/} {\bf 104} 073904

\bibitem{Brossarda}
 2011 {\em AIP Conf. Proc.\/} vol 1399

\bibitem{Lvovsky}
Lvovsky A~I, Sanders B~C and Tittel W 2009 {\em Nature (London)\/} {\bf 3}
  706--714

\bibitem{Gmitter}
Yablonovitch E, Gmitter T~J and Leung K~M 1991 {\em Phys.\ Rev.\ Lett.\/} {\bf
  67} 2295--2298

\bibitem{Ho}
Ho K, Chan C, Soukoulis C, Biswas∗ R and Sigalas M 1994 {\em Solid\ State\
  Commun.\/} {\bf 89} 413--416

\bibitem{KRAUSS}
Krauss T~F, Delarue R~M and Brand S 1996 {\em Nature\/} {\bf 383} 699--702

\bibitem{Steven}
Johnson S~G, Fan S, Villeneuve P~R, Joannopoulos J~D and Kolodziejski L~A 1999
  {\em Phys.\ Rev.\ B\/} {\bf 60} 5751--5758

\bibitem{Chow}
Lin S~Y, Chow E, Johnson S~G and Joannopoulos J~D 2000 {\em Opt.\ Lett.\/} {\bf
  25} 1297--1299

\bibitem{Villeneuve}
Chow E, Lin S, Johnson S, Villeneuve P, Joannopoulos J, Wendt J, Vawter G,
  Zubrzycki W, Hou H and Alleman A 2000 {\em Nature\/} {\bf 407} 983--986

\bibitem{Noda}
Chutinan A and Noda S 2000 {\em Phys.\ Rev.\ B\/} {\bf 62} 4488--4492

\bibitem{Nekuee}
Nekuee S~A~H, Akbari M and Mehrany K 2011 {\em IEEE Photonics Journal\/} {\bf
  3} 1111--1122

\bibitem{Shi}
Shi S, Chen C and Prather D~W 2004 {\em J.\ Opt.\ Soc.\ Am.\ A\/} {\bf 21}
  1769--1775

\bibitem{Qiu}
Qiu M 2002 {\em Appl.\ Phys.\ Lett.\/} {\bf 81} 1163--1165

\bibitem{Aram}
Aram M~H and Khorasani S 2015 Efficient analysis of photonic crystal slabs {J.
  Laser Opt Photonics} (accepted to be published)

\bibitem{Shi2}
Shi S, Chen C and Prather D~W 2005 {\em Appl.\ Phys.\ Lett.\/} {\bf 56} 043104

\end{thebibliography}

\end{document}